\documentclass[aps,10pt,twocolumn,noshowpacs,pre]{revtex4-1}
\usepackage{siunitx,latexsym,amssymb,epsfig,graphicx, amsmath,bm}
\usepackage{booktabs}
\usepackage{gensymb}

\begin{document}

\title{Corrugated silicon metasurface optimized within the Rayleigh hypothesis for anomalous refraction at large angles}

\author{Alexander A. Antonov and Maxim V. Gorkunov*}

\affiliation{Shubnikov Institute of Crystallography, FSRC ``Crystallography and Photonics'', Russian Academy of Sciences, 119333 Moscow, Russia\\
National Research Nuclear University MEPhI (Moscow Engineering Physics Institute), 115409 Moscow, Russia}

\email{gorkunov@crys.ras.ru}


\begin{abstract}
We optimize optical performance of metasurfaces based on periodically corrugated silicon layers by adjusting the Fourier coefficients of their surface profile. For smooth corrugations, we demonstrate an excellent  quantitative accuracy of semi-analytical approach based on the Rayleigh hypothesis. We employ the approach to design metasurfaces with anomalous refraction due to dominant first order diffraction. Unlike conventional Huygens' dielectric  metasurfaces, corrugated silicon layers are capable of efficient anomalous refraction in grazing directions: we obtain corrugation shapes allowing to deflect 70--80\% of the energy of normally incident green light into the range of $\ang{68}$--$\ang{85}$ of angles with respect to the normal.
\end{abstract}
\maketitle

\section{Introduction}\label{sec:intro}

The key advantage of metasurface concept is the possibility to precisely optimize the functionality by accurately designing the structural elements and adjusting their regular planar arrangement \cite{Yu2014}. While early realizations were based on arrays of metallic elements, structures made of high refractive index dielectrics are now obviously dominating the optical range \cite{Glybovski2016}. From the broad variety of formally suitable materials, silicon is the most popular one, as it combines a high refractive index with moderate optical losses \cite{Jellison1992, Vuye1993}, its samples of perfect quality are widely available and can be processed with nanofabrication techniques developed for nanoelectronics \cite{Jahani2016}. 

Accordingly, the majority of to date dielectric optical metasurfaces are based on thin silicon layers lithographically cut into arrays of nanoridges\cite{Sell2016}, nanorods\cite{Wu2014}, nanodisks \cite{Decker2015}, nanopillars \cite{Wang2016}, nanoposts \cite{Arbabi2015} or nanofins\cite{Khorasaninejad2016}, i.e., prisms having bases of various two-dimensional shapes and vertical walls of the height fixed by the initial silicon layer thickness. In this paradigm, the metasurface design is essentially reduced to the optimization of the base shape patterns. While ``brute force'' approaches relying on repetitive solution of the full-scale electrodynamic problem can be useful for gradual optimization \cite{Sell2017}, guidance by semi-analytical recipes greatly facilitates the progress. Thus presenting the metasurface response in terms of excitation of discrete sets of high quality Mie-type resonances hosted by dielectric prisms \cite{Kuznetsov2016} and exploiting Huygens' principle \cite{Decker2015} enabled predictable fabrication of metasurfaces for broadband highly efficient holography \cite{Wang2016}, nonlinear light deflection \cite{Wang2018} and even for parallel manipulation with multiphoton quantum states \cite{Wang2018a}.

All metasurface functionalities that require deflection of light propagation, eventually rely on the diffraction phenomenon: superwavelength complex unit cells are carefully constructed from  subwavelength elements (e.g. prisms) in a way that ensures domination of a particular diffraction order. When the efficiency of the latter is strong enough, the path of the energy flow becomes effectively broken and one speaks of anomalous refraction \cite{Yu2011}. As long as the angles at which the energy is deflected stay small, maximizing the diffraction efficiency does not pose a serious problem. However, simple Huygens' metasurface design fails when the refraction at larger angles is necessary and, especially, if a normally incident beam is to be deflected into a grazing outgoing direction \cite{Estakhri2016}. Bypassing the fundamental angular limitations can be achieved by introducing auxiliary leaky modes \cite{Diaz-Rubio2017,Asadchy2017} or bianisotropic elements \cite{Epstein2016}, which require precise engineering of the metasurface unit cells on deeply subwavelength scale and have been realized in the microwave and infrared ranges so far. For the near infrared, arrays of complex shaped silicon elements have been specifically designed and demonstrated an efficiency about 80\% of the refraction of light of 1050~nm wavelength from the normal incidence into the direction forming an angle of 75$^\circ$  with the surface normal \cite{Sell2017}. We are unaware of metasurfaces refracting visible light at such large angles.

As new nanotechnological approaches are being constantly introduced for the optical metasurface fabrication, the multitude of their designs steadily expands. For example, silicon nanostructures of truly three-dimensional chiral shapes have been lately imprinted in monocrystalline silicon using digitally controlled focused ion beam \cite{ROGOV2017}. The technique is capable of producing smooth complex shaped periodic corrugations of the silicon surface, and allowed creating  highly transparent metasurfaces with strong optical chirality \cite{Gorkunov2018}. While numerical modeling has linked the peculiar optical properties with high-quality dielectric resonances, approaches for the precise design of such metasurfaces are yet to be developed. 

In this paper, we propose a systematic method of optimizing corrugated dielectric metasurfaces and demonstrate its potential on the simplest case of one-dimensionally corrugated silicon layers. We represent the surface profile in terms of a few lowest Fourier harmonics and reduce the task to the search for several Fourier coefficients. As is described in Section~\ref{sec:analytical}, applying the Rayleigh hypothesis (RH) yields accurate semi-analytical solution of the light transmission problem, which drastically  facilitates the optimization. Choosing the efficiency of diffraction into the +1 channel at a given wavelength as a particular figure of merit, we find in Section~\ref{sec:optimiztion} the corresponding optimal surface profiles paying special attention to those with the period sightly larger than the operational wavelength. The mechanism and practical prospects of strong anomalous refraction of visible light into grazing directions are discussed in Section~\ref{sec:discuss}. The conclusions are summarized in Section~\ref{sec:concl}.

\begin{figure}
	\centering\includegraphics[width=\columnwidth]{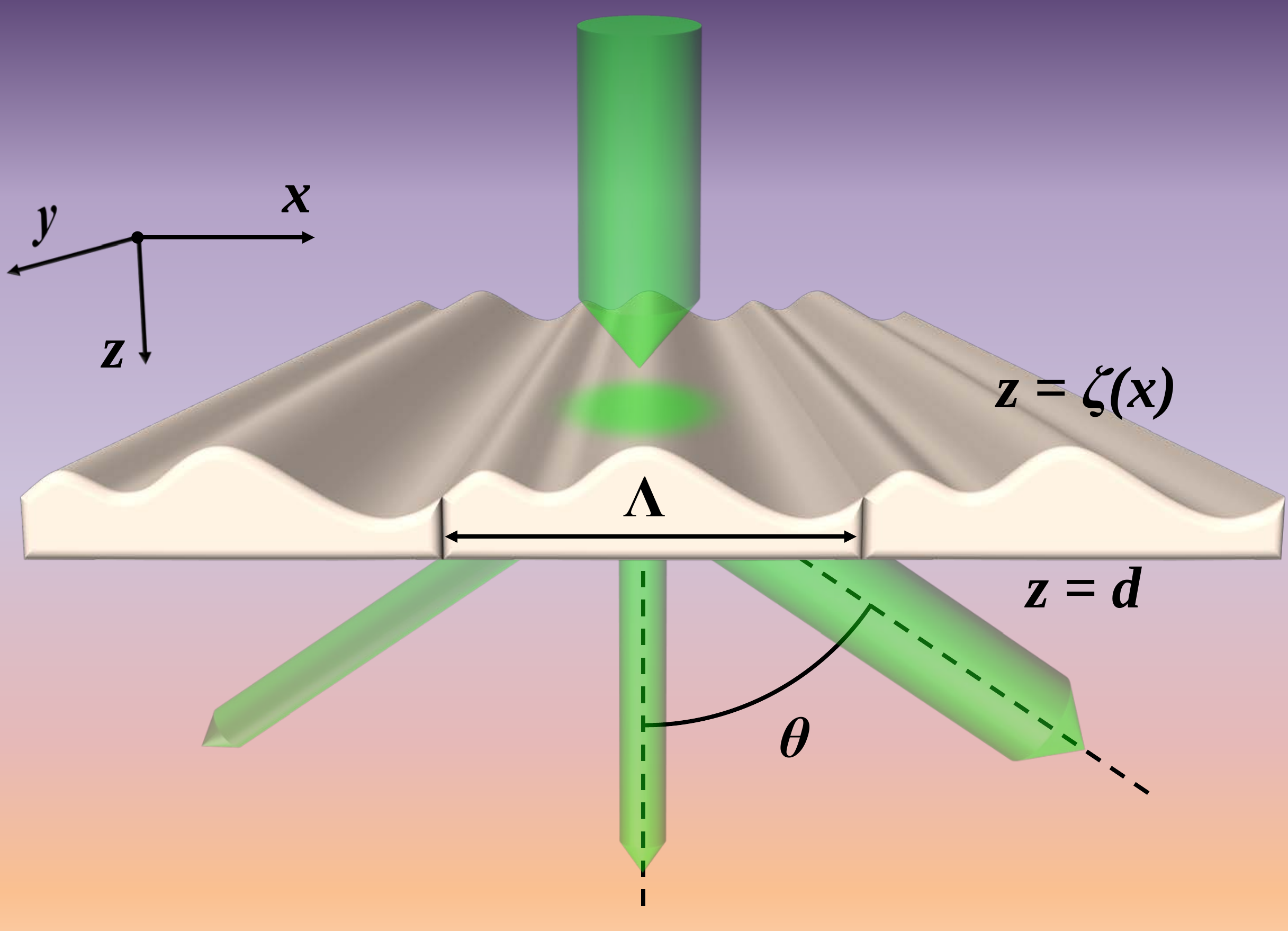}
	\caption{Artistic view of the problem of light transmission and diffraction by a corrugated silicon metasurface.}\label{fig:scheme}
\end{figure}

\section{Semi-analytical solution in terms of the Rayleigh hypothesis}\label{sec:analytical}
\subsection{Basic relations}

Consider normal incidence of a TM-polarized monochromatic plane wave onto a layer of dielectric material of permittivity $\varepsilon$ confined between the plane $z=d$ and the surface corrugated along the profile $z=\zeta(x)$ with a period $\Lambda$, see Fig.~\ref{fig:scheme}. Everywhere in this paper we assume $\varepsilon$ to take the complex values of silicon permittivity measured at $\ang{20}$C \cite{Vuye1993}.

For such case, RH formally presumes the existence of only one primary wave incident on any point of the $\zeta(x)$ surface, while secondary waves emitted by one part of this surface and incident on the other are neglected. Upon its first introduction in 1907 for reflective perfect metal gratings~\cite{Rayleigh1907}, RH has been recognized to yield semi-analytical solutions of electrodynamic problems involving periodically corrugated interfaces, while its validity is being severely debated ever since. Reasonable arguments relating the accuracy of RH with the interface shape \cite{Millar1971} and smoothness \cite{Berg1979} have been formulated. Later, an ambiguity of the near-field separation into incoming and outgoing waves has been pointed out, which explained the surprising quantitative accuracy and large range of applicability of this seemingly very rough approximation \cite{Voronovich2007}. Directly comparing semi-analytical solutions with those obtained by accurate numerics indicated also an important role of errors accumulated during integration and operation with ill conditioned matrices  \cite{Tishchenko2009}. Being aware of all the uncertainties, in the following we repeatedly verify our semi-analytical results against direct numerical solutions of the Maxwell equations.

Assuming that all monochromatic fields depend on time as $e^{-i\omega t}$ and taking for simplicity a unit incident field amplitude, we apply RH and write the $y$-component of the complex magnetic field amplitude as:
\begin{multline}\label{Hy}
H_{y}(x,z) =\\
\left\{\begin{array}{l}e^{ik_{0}z} + \sum_{m}a_{1m}e^{imKx - ik_{m}z},\  z<\zeta(x)\\\sum_{m}\left(a^{+}_{2m}e^{i\kappa_{m}z} + a^{-}_{2m}e^{- i\kappa_{m}z}\right)e^{imKx}, \ \zeta(x)<z<d\\\sum_{m}a_{3m}e^{imKx + ik_{m}(z-d)},\ \ z>d\end{array}\right.
\end{multline}
where $K = {2\pi}/{\Lambda}$ is the corrugation wavenumber, while
\begin{equation}
k_{m} = \sqrt{\left({2\pi}/{\lambda}\right)^2 - \left(mK\right)^2},\ \kappa_{m} = \sqrt{\varepsilon\left({2\pi}/{\lambda}\right)^2 - (mK)^2}\\.
\end{equation} 
The field \eqref{Hy} has to be continuous across both dielectric interfaces together with its derivative along the local surface normal divided by the permittivity, ${{\varepsilon}^{-1}\partial H_{y}}/{\partial n}$.

Following the routine from Ref.~\cite{Tishchenko2009}, to relate the field amplitudes below and above the corrugated surface, we equate the fields on its both sides, multiply them by $e^{-iqKx}$ with an integer $q$, and integrate over one period $\Lambda$. This yields a set of equations:
\begin{equation}\label{set1}
I^{1+}_{q0} + \sum_{m}a_{1m}I^{1-}_{qm}= \sum_{m}\left(a_{2m}^{+}I^{2+}_{qm}+a_{2m}^{-}I^{2-}_{qm}\right),
\end{equation}
where the coefficients are expressed as integrals:
\begin{equation}
\begin{array}{c}\label{Iqm}
{I}^{1\pm}_{qm} = \frac{1}{\Lambda}\underset{0}{\overset{\Lambda}{\int }}\exp\left[iK(m-q)x{\pm}i{k}_{m}\zeta(x)\right]dx\\ {I}^{2\pm}_{qm} = \frac{1}{\Lambda}\underset{0}{\overset{\Lambda}{\int }}\exp\left[iK(m-q)x{\pm}i{\kappa}_{m}\zeta(x)\right]dx
\end{array}
\end{equation}

\begin{figure*}
	\centering\includegraphics[width=\textwidth]{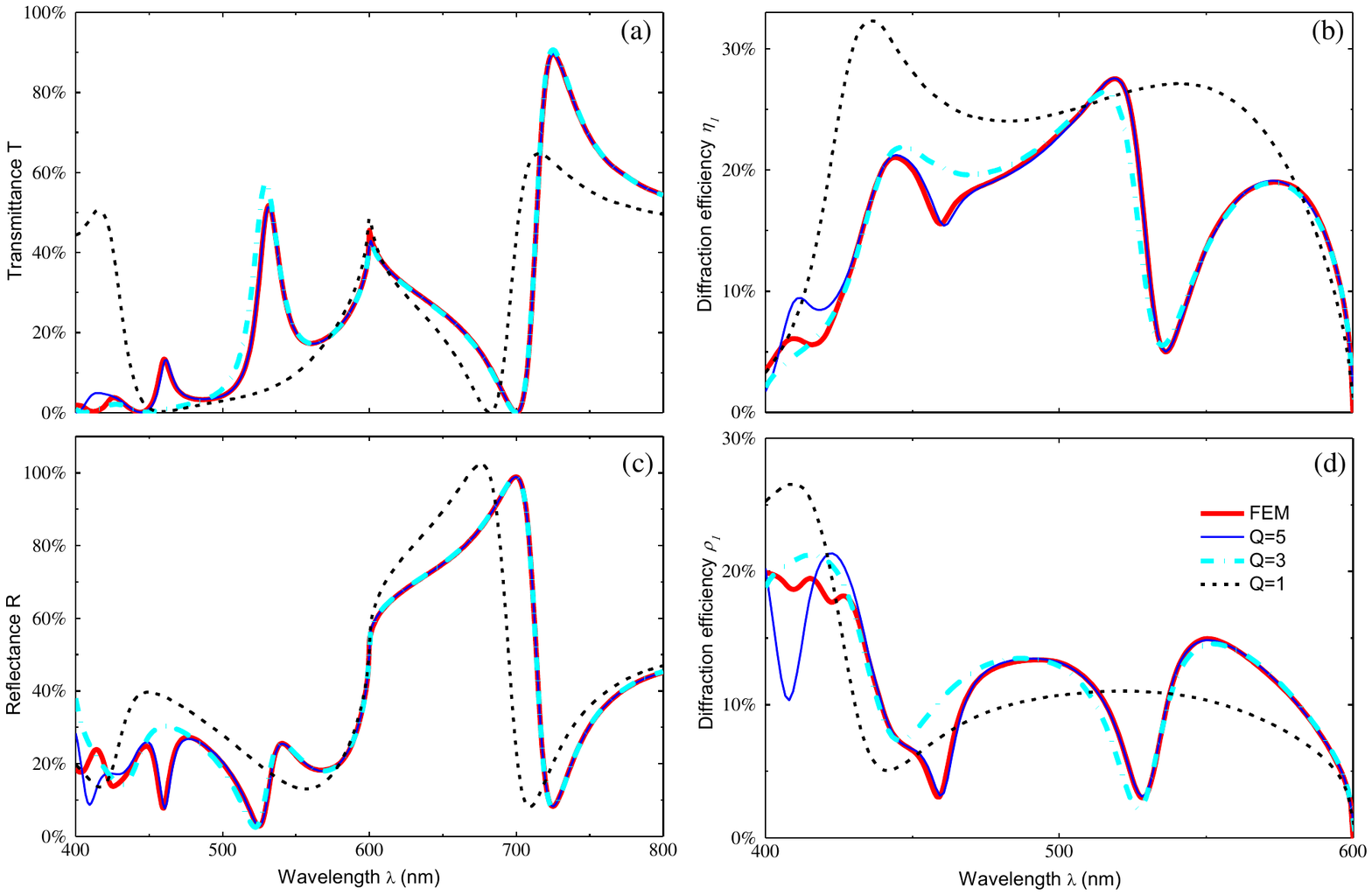}
	\caption{Optical properties of corrugated silicon layer confined between the surfaces $z=d$ and $z = \Delta\cos({2\pi x}/{\Lambda})$ with $d = 50$~nm, $\Delta = 45$~nm and  $\Lambda = 600$~nm. The spectra of transmittance (a), reflectance (b), and transmission and reflection diffraction efficiencies (c) and (d) calculated within RH for different truncation numbers $Q$ are plotted together with those obtained by FEM numerical modelling (as indicated on the legend in (d)).}\label{fig:cos}
\end{figure*}

For a differentiable corrugation profile with finite $\zeta'(x)=d\zeta/dx$, the field derivative along the surface normal  ${\partial H_{y}}/{\partial n}$ satisfies the relation:
\begin{equation}
\sqrt{1+{[{\zeta}^{\prime }(x)]}^{2}}\frac{\partial H_{y}}{\partial n} = \frac{\partial H_{y}}{\partial z} - {\zeta}^{\prime }(x)\frac{\partial H_{y}}{\partial x},
\end{equation}
and the corresponding continuity condition can be applied directly to the right-hand side here. Taking the  derivatives of fields from Eq.~\eqref{Hy}, multiplying them by $e^{-iqKx}$ and integrating by parts over one grating period yields another set of equations relating the amplitudes $a_{1m}$ and $a_{2m}^\pm$:
\begin{multline}\label{set2}
\varepsilon k_{0}I^{1+}_{q0} - \sum_{m}a_{1m}\varepsilon I^{1-}_{qm}\left({k^{2}_{0}-mqK^{2}}\right)k_{m}^{-1}=\\ \sum_{m}\left(a_{2m}^{+}I^{2+}_{qm}-a_{2m}^{-}I^{2-}_{qm}\right)\left(\kappa^{2}_{0}-mqK^{2}\right)\kappa_{m}^{-1}
\end{multline}

For the flat $z=d$ interface, the continuity conditions relate the field amplitudes below and above it in a standard way, yielding another two trivial sets of equations for the amplitudes $a_{3m}$ and $a_{2m}^\pm$. Combining them with the sets \eqref{set1} and \eqref{set2} we exclude the amplitudes $a_{2m}^\pm$ and obtain closed sets of equations relating the amplitudes of the fields above and below the metasurface:
\begin{multline}\label{set3}
{I}^{1+}_{q0} + \sum_{m}a_{1m}{I}^{1-}_{qm} =\sum_{m}a_{3m}\left(\psi^{+}_m{I}^{2+}_{qm} + \psi^{-}_m{I}^{2-}_{qm}\right),
\end{multline}
\begin{multline}
\varepsilon k_{0}{I}^{1+}_{q0} -\sum_{m}a_{1m}\varepsilon{I}^{1-}_{qm}\ \left({k^{2}_{0}-mqK^{2}}\right)k_{m}^{-1}=\\ \sum_{m}a_{3m}\left(\psi^{+}_m{I}^{2+}_{qm} - \psi^{-}_m{I}^{2-}_{qm}\right)\left(\kappa^{2}_{0}-mqK^{2}\right)\kappa_{m}^{-1}.\label{set4}
\end{multline} 
where $\psi^{\pm}_m = {e^{\mp i\kappa_{m}d}}\left(1\pm {k_{m}\varepsilon\kappa_{m}^{-1}}\right)/2$.

Obviously, the sets \eqref{set3} and \eqref{set4} are to be truncated in order to be solved. We introduce a truncation number $Q$ and consider equations with integer $q=-Q,...,Q$ and for the field harmonic amplitudes with integer $m=-Q,...,Q$.
Solving the set of remaining $(4Q+2)$ linear algebraic equations, we evaluate the key optical observables, such as the transmittance $T = |a_{30}|^{2}$, the reflectance $R = |a_{10}|^{2}$, as well as the  efficiencies of diffraction into $m$-th transmitted order $\eta_{m} = |a_{3m}|^{2}\sqrt{1-\left({m\lambda}/{\Lambda}\right)^{2}}$ and $m$-th reflected order $\rho_{m} = |a_{1m}|^{2}\sqrt{1-\left({m\lambda}/{\Lambda}\right)^{2}}$.

%

\subsection{Numerical validation}

To verify the results based on RH, we employ \textsc{Comsol Multiphysics} software using the  electromagnetic wave frequency domain (ewfd) solver of the Wave Optics module based on the finite elements method (FEM). The two-dimensional problem is solved in the domain $0\le x\le\Lambda$,  -1200~nm$\le z\le$1200~nm with periodic boundary conditions along the $x$-axis and perfectly matched absorbing layers at the top and the bottom. The monochromatic wave source and line detectors are positioned at $z=-1200$~nm and $z=1200$~nm correspondingly. The mesh element size is kept between 0.3~nm and 9~nm in silicon and between 1.5~nm and 50~nm in the air. 

For clarity, we separate from the effects of corrugation shape by considering the simplest single-cosine corrugation and compare the optical observables obtained with different truncation numbers $Q$ and with \textsc{Comsol Multiphysics} numerical solver. As shown in Fig.~\ref{fig:cos}, increasing $Q$ to rather moderate values allows reproducing the accurate numerical solution in great detail. Remarkably, accounting already for a few first field harmonics provides a good estimate of the spectra in general. Increasing $Q$ to as little as 5 we manage to reproduce finer acute features and also achieve excellent quantitative accuracy across the most part of the visible range. Note, however, that resolving certain weak  features in the blue range requires higher $Q$. 

To understand such behavior, we remind of the physical nature of spectral anomalies which can be understood in terms of Wood anomalies and guided mode resonances. The former occur at the cut-off wavelengths of diffraction orders. Here a cut-off of the first diffraction order takes place at a wavelength of 600~nm . As seen in Fig.~\ref{fig:cos} the Wood anomaly there is nicely reproduced already with $Q=1$, i.e., by taking into account the involved $\pm1$ field harmonics.

The guided mode resonances appear when the surface corrugations couple the modes guided by the layer of silicon to the incident and outgoing free-space plane waves. 
Stronger resonances appear due to stronger first-order coupling, and a pair of such resonances manifest themselves as dips of the transmittance down to almost zero values around 450~nm and 700~nm wavelengths. Predictably, these resonances are also nicely reproduced with only a few modes taken into account. According to the accurate numerical solution, there are also several weaker resonances hosted by the cosine corrugated layer. Increasing $Q$ introduces into consideration higher field harmonics coupled to those weaker resonances, and the corresponding  finer dips and peaks on the spectra become also resolved. 

It would be a mistake, however, to recommend increasing $Q$ up to much larger values. While large $q$ require evaluating integrals of very fast oscillating exponents in Eqs.~\eqref{Iqm}, introducing high-order weakly excited field harmonics also makes the matrix of the corresponding system of linear equations ill defined. As discussed in Ref.~\cite{Tishchenko2009}, this results in critical accumulation of numerical errors and unpredictably diverging solutions. 

\begin{figure}
	\centering\includegraphics[width=0.95\columnwidth]{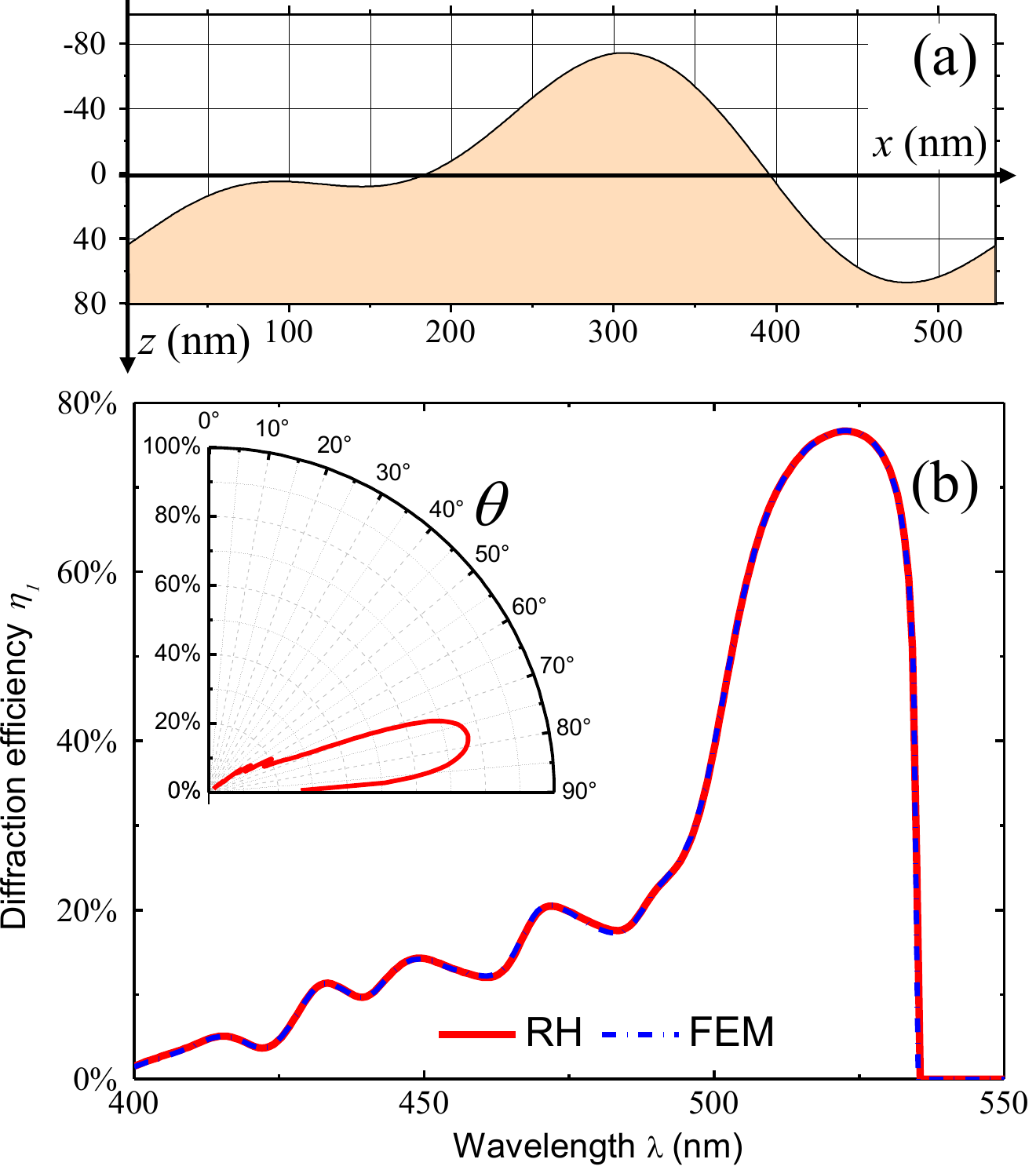}
	\caption{Corrugated metasurface $N=2$ of a period of 535~nm optimized for the maximum +1 transmitted diffraction at a 532~nm wavelength: (a) cross section of a metasurface unit cell; (b) spectra of diffraction efficiency $\eta_{1}$ obtained within RH with $Q=12$ (solid line) and with FEM by \textsc{Comsol} (dashed line). The inset in (b) illustrates how $\eta_{1}$ is distributed over the diffraction directions characterized by the angle $\theta$ to the surface normal.}\label{fig:N2}
\end{figure}

%
%
%
%

\section{Optimization for anomalous refraction}\label{sec:optimiztion}

\begin{table*}
	\centering
	\begin{ruledtabular}
		\begin{tabular}{ccccccccccc}
			\toprule
			Number & \multicolumn{10}{c}{Parameters (nm)} \\ \cmidrule(l){2-11}
			of harmonics & $c_{1}$ & $c_{2}$ & $s_{2}$ & $c_{3}$ & $s_{3}$ & $c_{4}$ & $s_{4}$ & $\Delta$ & $d$ & $\Lambda$ \\
			\hline
			\\
			$N=2$ & 52.09 & -7.69 & -28.78 & - & - & - & - & 60 & 80 & 535 \\
			$N=3$ & 50.57 & -14.10 & -27.34 & -8.77 & 4.45 & - & - & 60 & 80 & 535 \\
			$N=4$ & 48.95 & -12.74 & -30.08 & -11.05 & 3.83 & -5.01 & 8.20 & 60 & 80 & 535\\    
			\\
			\bottomrule
		\end{tabular}
		\caption{Parameters of the optimized silicon metasurfaces with different number of corrugation harmonics $N$.}\label{tab}
	\end{ruledtabular}
\end{table*}

The possibility to obtain reliable spectra of optical observables by solving the sets of linear algebraic equations \eqref{set3} and \eqref{set4} greatly enhances the speed and efficiency of the  optimization of corrugated metasurfaces. Smoothness of the surface corrugation in available experimental samples \cite{Gorkunov2018} as well as the RH applicability range both suggest focusing on smooth profiles $\zeta(x)$. Accordingly, we present the latter as Fourier series expecting a few first terms to be of prime importance:
\begin{equation}\label{zetaF}
\zeta(x) = \sum^{N}_{n=1}\left[c_{n}\cos(nKx)+s_{n}\sin(nKx)\right].
\end{equation}

By setting $s_1=0$ we eliminate the ambiguity of the $x$-axis origin. To stabilize the numerical optimization against uncontrollable growth of the profile amplitude, we limit the maximum depth of corrugation \eqref{zetaF} by fixing the parameter $\Delta^2={\sum_{n=1}^{N}(c_{n}^{2}+s_{n}^{2})}$. Under this restriction, the Fourier coefficients can be parametrized by the sets of angles $\{\alpha_2,...,\alpha_N\}$ and $\{\beta_2,...,\beta_N\}$:
\begin{eqnarray*}
	&c_{1} = \Delta\cos\beta_{2}\cos\beta_{3}\cos\beta_{4}...\cos\beta_{N},\\
	&c_{2} = \Delta\cos\alpha_{2}\sin\beta_{2}\cos\beta_{3}\cos\beta_{4}...\cos\beta_{N} \\	
	&s_{2} = \Delta\sin\alpha_{2}\sin\beta_{2}\cos\beta_{3}\cos\beta_{4}...\cos\beta_{N} \\
	&c_{3} = \Delta\cos\alpha_{3}\sin\beta_{3}\cos\beta_{4}...\cos\beta_{N} \\	
	&s_{3} = \Delta\sin\alpha_{3}\sin\beta_{3}\cos\beta_{4}...\cos\beta_{N} \\		
	&...\\
	&c_{N} = \Delta\cos\alpha_{N}\sin\beta_{N}\\
	&s_{N} = \Delta\sin\alpha_{N}\sin\beta_{N}
\end{eqnarray*}
which we consider as independent optimization variables. 

We use \textsc{Matlab} fminsearch routine to maximize the efficiency $\eta_{1}$ of the +1 order of transmitted diffraction at a standard green light wavelength of $\lambda = 532$~nm. The goal is to obtain a metasurface redirecting considerable amount of the energy of normally incident light into oblique direction as shown in Fig.~\ref{fig:scheme}. We set the silicon layer thickness to $d = 80$~nm,  the modulation depth being kept as $\Delta = 60$~nm, while the period is fixed to $\Lambda=535$~nm. Note that the closeness of $\Lambda$ to $\lambda$ determines a grazing angle $\theta\approx\ang{83.9}$ of the maximized first order diffraction.
 
\begin{figure}
	\centering\includegraphics[width=0.95\columnwidth]{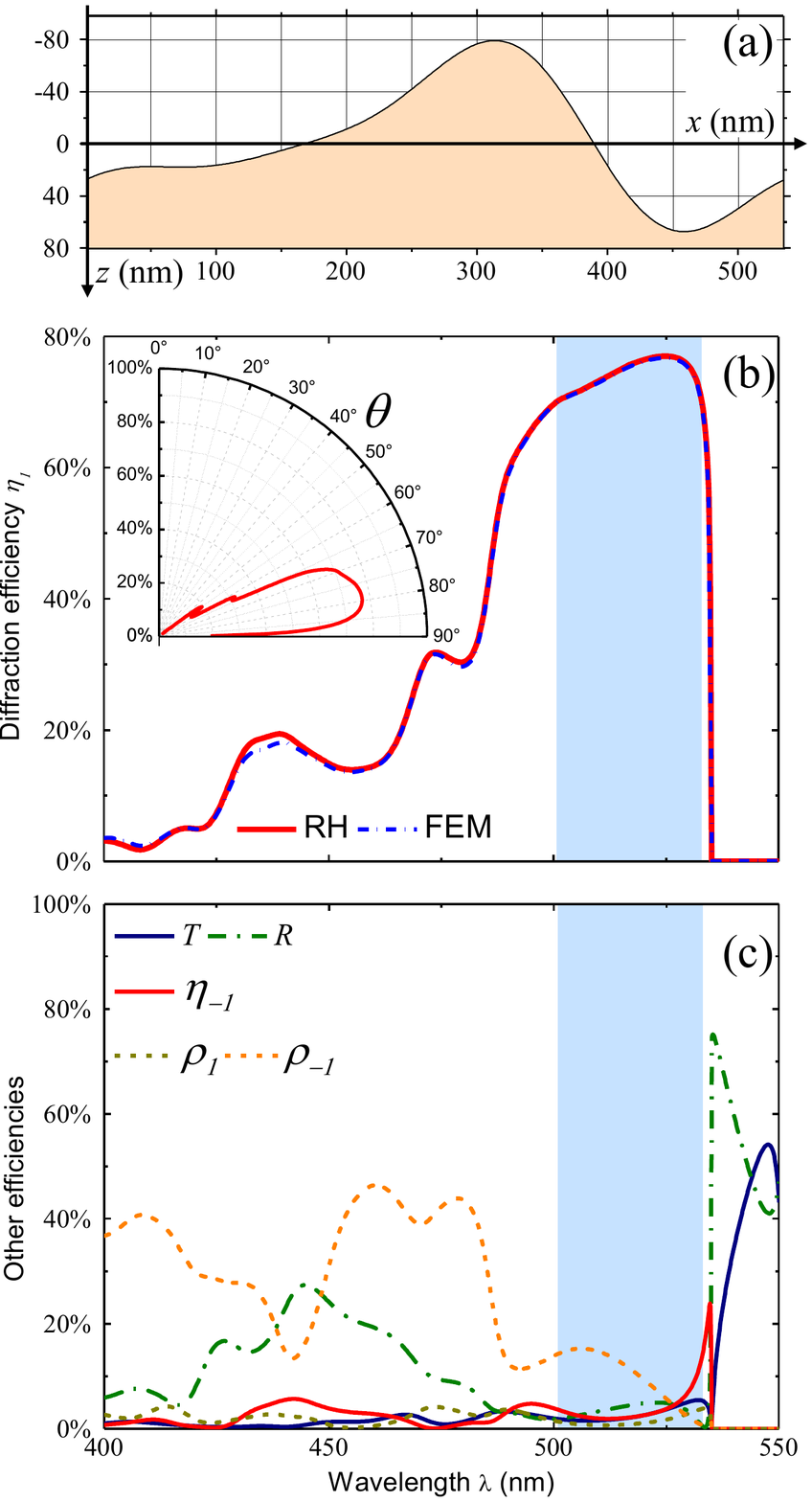}
	\caption{Corrugated metasurface of a period of 535~nm optimized with $N=3$ for the maximum +1 transmitted diffraction at a 532~nm wavelength: (a) cross section of a metasurface unit cell; (b) spectra of diffraction efficiency $\eta_{1}$ obtained within RH with $Q=14$ (solid line) and with FEM by \textsc{Comsol} (dashed line); (c) other channels spectra: transmittance~$T$, reflecttance~$R$, transmitted diffraction efficiency $\eta_{-1}$ and reflected diffraction efficiencies $\rho_{\pm1}$. The inset in (b) illustrates how $\eta_{1}$ is distributed over the diffraction directions characterized by the angle $\theta$  from the surface normal. Background stripes in (b) and (c) highlight the range of diffraction efficiency  $\eta_{1}\ge70\%$}\label{fig:N3}
\end{figure}

As seen in Fig.~\ref{fig:cos}, the diffraction spectra are rather complex already for the silicon layer with the single-cosine corrugation. As one can hardly expect their considerable simplification for corrugations with several Fourier harmonics, the success of optimization critically depends on the choice of initial values of the variables $\{\alpha_2,...,\alpha_N\}$ and $\{\beta_2,...,\beta_N\}$. This problem can be solved in a straight forward manner for double-periodic corrugations with $N=2$, by exhausting the possibilities in a grid-search.  
The calculation speed granted by RH, allows us to check all sets of initial $\alpha_{2}$ and $\beta_{2}$ values from the intervals $[0; 2\pi]$ with a $\pi/20$ step for the truncation number set to $Q=10$. In the most cases, the optimization safely converges to the same surface profile shown in Fig.~\ref{fig:N2}a). The particular optimal corrugation parameters are listed in Table~\ref{tab}. As seen in Fig.~\ref{fig:N2}b),  the diffraction efficiency exceeds the 70\% level within a narrow spectral rage, which, as is seen in the inset, corresponds to the diffraction angles between $\ang{72}$ and $\ang{83}$. A quantitative comparison with the solution by FEM demonstrates that an accuracy  of 1.5\% is achieved in the whole spectral range with RH-based solution with $Q=10$. Increasing the truncation number to $Q=12$ allows obtaining the spectra practically identical to those resolved by FEM (see Fig.~\ref{fig:N2}b).

For more complex surface profiles, we are unable to perform a full grid-search for the best set of initial values. Instead, we proceed to $N=3$, taking for the initial values of  $\alpha_{2}$ and $\beta_{2}$ the optimal ones obtained above, and perform a grid search for the initial values of parameters $\alpha_{3}$ and $\beta_{3}$ again from $[0; 2\pi]$ intervals with a $\pi/20$ step. To minimize the errors, we set $Q=12$. Again, most of successful optimization tries converge to a well defined optimal configuration. The result is presented in Fig.~\ref{fig:N3}, while the optimal set of parameters can be found in Table~\ref{tab}. The obtained optical observables are also verified against those obtained by FEM and the deviations are truly negligible.

\begin{figure}
	\centering\includegraphics[width=\columnwidth]{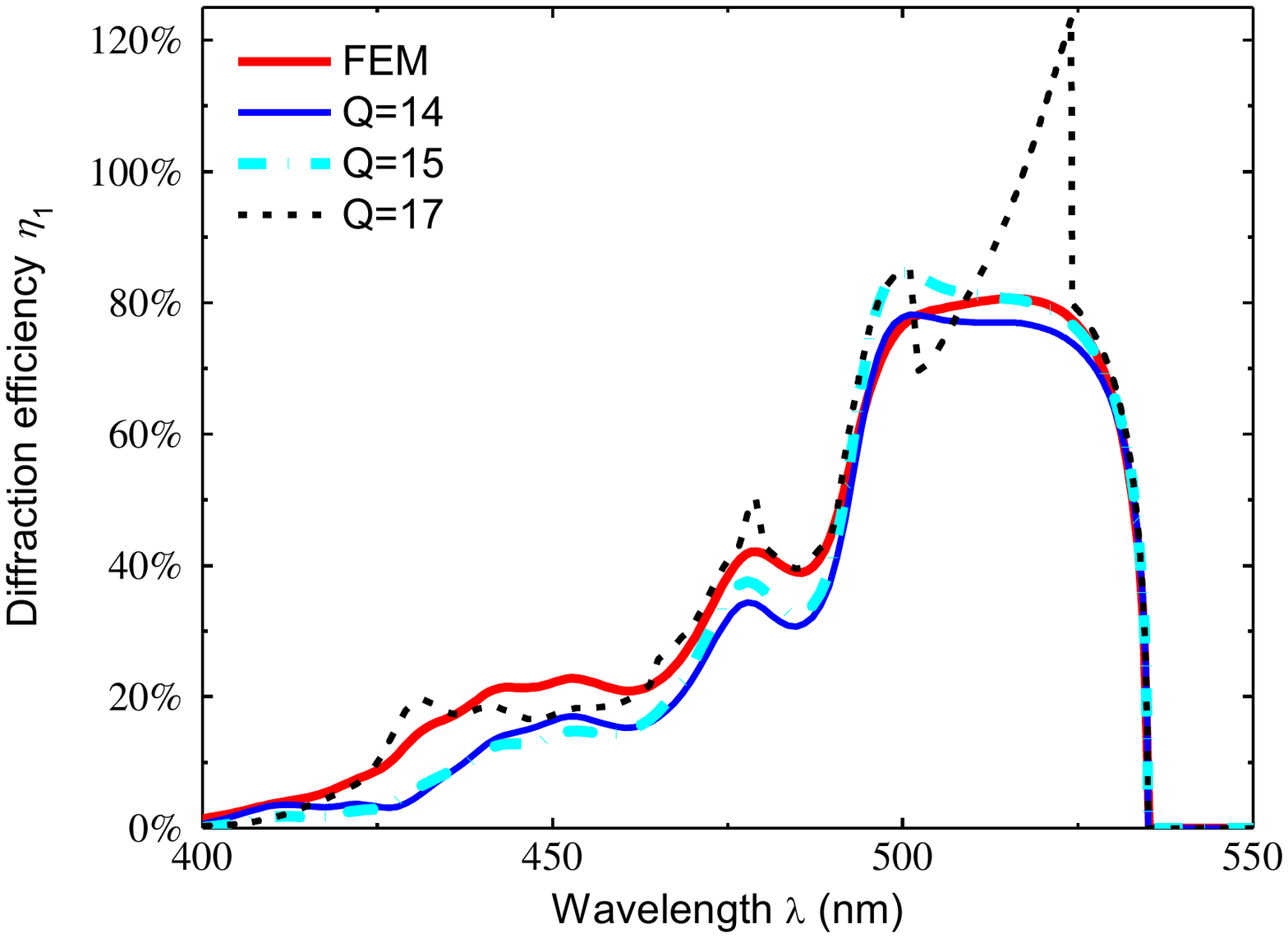}
	\caption{Spectra of diffraction efficiency of corrugated metasurface of a period of 535~nm optimized with $N=4$ for the maximum +1 transmitted diffraction at a 532~nm wavelength obtained within RH with different $Q$ and with FEM by \textsc{Comsol} as indicated in the legend.}\label{fig:N4}
\end{figure}

One can see that adding higher harmonics leaves the maximum level of $\eta_1$ practically unaffected. At the same time, it substantially broadens the spectral range of strong first order diffraction: the values of diffraction efficiency exceeding 70\% are now attainable starting from the wavelength of 500 nm. The corresponding angles of diffraction here cover the range from $\ang{68}$ to $\ang{85}$.  Note, that in this range (highlighted by a blue stripe in Fig.~\ref{fig:N3}), the efficiencies of other output channels do not exceed 15\%, the strongest being the reflection diffraction with efficiency $\rho_{-1}$, which is directly counterpropagating  to the maximized channel with $\eta_1$.

Following the same strategy, we proceed to optimizing the corrugation with $N=4$. The obtained structure parameters are listed in Table~\ref{tab}. Plots of the corresponding spectra of the diffraction efficiency in Fig.~\ref{fig:N4} obtained with several close truncation numbers $Q$ appear to be poorly converging to the  accurate FEM-resolved solution. Increasing $Q$ to a rather moderate value of $Q=17$ we obtain a strong divergence: $\eta_1$ exceeds by far even its natural limit of 100\%, which we attribute to numerical errors accumulated during the integrations and ill defined matrix operations. Such type of behavior is well known \cite{Tishchenko2009}, and it indicates crossing the limits of RH applicability. At the same time, comparing the FEM-resolved spectra of $\eta_1$ in Figs.~\ref{fig:N3}b) and \ref{fig:N4} demonstrates that adding another harmonic allows further broadening of the anomalous refraction band.

\section{Discussion}\label{sec:discuss}

As we have shown, it is possible to realize efficient anomalous refraction of green light into grazing directions with silicon layers having rather simple smooth periodic surface profiles. 
Visually comparing the cross sections shown in Figs.~\ref{fig:N2}a)  and \ref{fig:N3}a) one can conclude that the optical performance of such metasurfaces is notably sensitive to the deviations of cross section shape. Quantitatively this can be estimated from comparing the amplitudes of corrugation harmonics presented in Table~\ref{tab}: all metasurfaces posses similar strong amplitudes $c_1$ of the fundamental harmonic; the amplitudes of substantial double-periodic modulation also differ only by a few nanometers; while higher-order harmonics are either absent or relatively weak. Nevertheless, such subtle differences produce noticeable effect on the anomalous refraction range. From the practical point, this means that fabrication of corrugated metasurfaces requires technologies capable of precisely sustaining the corrugation profile, as deviations of its depth by about 10~nm noticeably affect the performance. At the same time, the in-plane resolution can be moderate: the characteristic sizes of necessary shape features are of the order of 100~nm. 

\begin{figure}
	\centering\includegraphics[width=0.7\columnwidth]{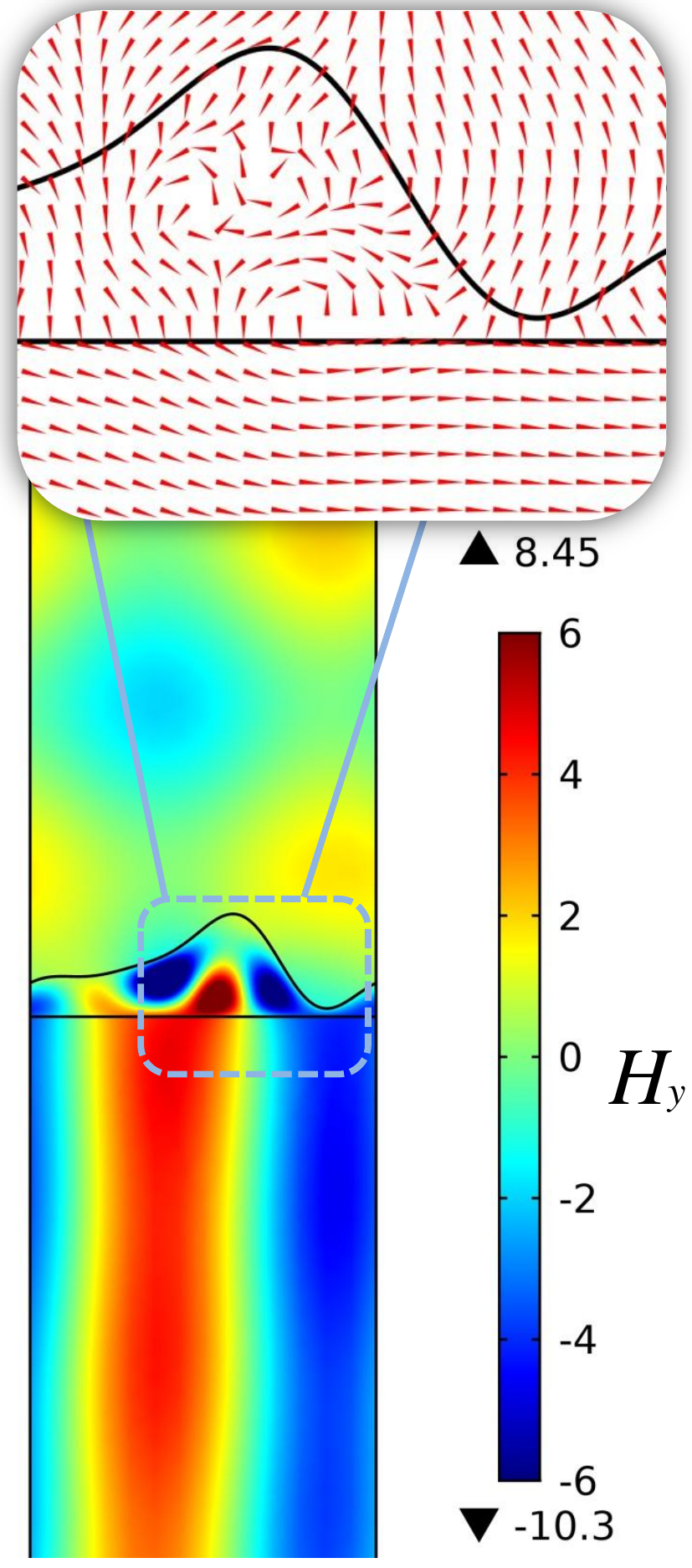}
	\caption{Distribution of magnetic field component $H_y$ inside and close to the 535~nm periodic metasurface optimized with $N=3$ (see the parameters in Table~\ref{tab}) as TM-polarized light of a 533.6~nm wavelength and unit amplitude is incident on it normally from the top and 66\% of the energy is refracted at an angle $\theta=\ang{86}$. The inset shows a local distribution of the Poynting vector direction.}\label{fig:fields}
\end{figure}

For a plausible explanation of the physics behind the strong anomalous refraction, we plot in Fig.~\ref{fig:fields} the spatial distribution of the magnetic field resolved by accurate \textsc{Comsol Multiphysics} modeling of an extreme case, when 66\% of the incident energy is refracted into a grazing direction only $\ang{4}$ away from the metasurface plane. As all efficiencies are defined according to the energy transported across a horizontal plane, 
the refracted wave amplitude exceeds here that of the incident wave by almost an order of magnitude, which makes the field pattern so contrast below the metasurface. Also above it, for the same reason, the reflected diffracted waves transporting very little energy in the vertical direction interfere so strongly with the incident wave of a unit amplitude. The most peculiar phenomena, however, occur inside the silicon layer, where one can see pronounced standing waves corresponding to a couple of dielectric Mie-type resonances. We presume the anomalous refraction to be determined by an interplay between those resonances, with the relative   phase and resonant wavelengths being determined by the cross section shape. In this context, our optimization can be understood as their careful adjustment via subtle shape variations. 

Note that the maximal thickness of the proposed metasurfaces is below 160~nm, i.e., they are of formally subwavelength thickness compared to the operational green light wavelength. 
It is interesting to consider how such corrugated silicon layers bypass the restrictions on anomalous refraction at grazing angles formulated in general for subwavelength thin metasurfaces (see e.g. Refs. \cite{Estakhri2016, Asadchy2016}). First, in spite of the small thickness, as is obvious from the field distribution in Fig.~\ref{fig:fields}, the metasurface response is substantially nonlocal and it cannot be reduced to a trivial effective boundary with inhomogeneous surface impedance. Next, considering the Poynting vector pattern shown as the inset in Fig.~\ref{fig:fields}, one notices a peculiar inflow and drain of electromagnetic energy in certain parts of the bottom flat silicon interface. In an effective boundary with local response, such alternating energy flow pattern is possible only when the surface contains interchanging regions with energy gain and loss. With passive metasurfaces, such patterns are typical of the structures guiding leaky modes that can absorb the electromagnetic energy at one point and release it at the other \cite{Asadchy2017}. Apparently, here we have a similar situation: the electromagnetic energy peculiarly circulates while being trapped inside the silicon layer.

Finally, our choice of operational wavelength of green light is totally arbitrary. Although scaling the metasurfaces down to the blue and ultraviolet ranges can be problematic as a substantial part of the light energy will be absorbed in silicon, scaling them up to longer wavelengths will hardly be a problem, as even better optical properties of silicon are then to be employed. Slight structure adjustments will be necessary though to compensate for the moderate frequency dispersion of silicon permittivity. 

\section{Conclusions}\label{sec:concl}
We propose a metasurface design based on periodic smooth corrugation of subwavelength thin dielectric layers. Optimization of such metasurfaces can be critically simplified if the Rayleigh hypothesis is applicable. We obtain particular exemplary silicon  metasurfaces that are capable of efficient visible light refraction into grazing directions.


\section*{Acknowledgments} 
The research is supported by the Russian Science Foundation (project 18-12-00361). The authors are grateful to Alexey Kondratov for the kind assistance with \textsc{Comsol Multiphysics} modelling.

%

\end{document}